\documentclass[a4paper,11pt]{article}
\usepackage{jheppub} 
\usepackage{lineno}
\usepackage[compat=1.0.0]{tikz-feynman}
\usepackage{float}
\usepackage{feynmp-auto}

\usepackage[utf8]{inputenc}
\usepackage[T1]{fontenc}
\usepackage{textcomp}
\usepackage{newunicodechar}

\newunicodechar{σ}{$\sigma$}
\newunicodechar{γ}{$\gamma$}
\newunicodechar{²}{$^2$}
\newunicodechar{ₜ}{$_t$}  
\newunicodechar{ₒ}{$_o$}  
\newunicodechar{ₙ}{$_n$}  
\newunicodechar{≳}{$\geq$}

\newcommand{\marrow}[5]{%
    \fmfcmd{style_def marrow#1
    expr p = drawarrow subpath (1/4, 3/4) of p shifted 6 #2 withpen pencircle scaled 0.4;
    label.#3(btex #4 etex, point 0.5 of p shifted 6 #2);
    enddef;}
    \fmf{marrow#1,tension=0}{#5}}

\arxivnumber{2509.11105} 

\title{\boldmath Focused Angular $N$-Body Event Generator (FANG)}

\author[a,b]{Itay Horin}
\affiliation[a]{Racah Institute of Physics, The Hebrew University of Jerusalem, 9190401, Israel}
\affiliation[b]{Soreq Nuclear Research Center, Yavne 81800, Israel}
\author[b]{Arik Kreisel}

\author[c]{Or Alon}
\affiliation[c]{Weizmann Institute of Science, Rehovot 7610001, Israel}

\emailAdd{itay.horin@mail.huji.ac.il}

\abstract{We introduce \textsc{FANG} (Focused Angular $N$-body event Generator), a new Monte Carlo tool for efficient event generation in restricted Lorentz-Invariant Phase Space (LIPS). Unlike conventional approaches that sample the full $4\pi$ solid angle, \textsc{FANG} directly generates events in which selected final-state particles are constrained to fixed directions or finite angular regions in the laboratory frame. Because of the way the generator is constructed, angular constraints can be imposed directly in the laboratory frame while maintaining the correct LIPS structure, enabling differential and total cross sections or decay rates to be computed with high efficiency. The method is validated against analytic results and existing event generators, showing excellent agreement.
By reducing the computational cost of full phase-space event generation by several orders of magnitude, \textsc{FANG} provides a robust and versatile framework applicable to particle, nuclear, and detector physics.}

\begin{document}
\maketitle
\flushbottom

\section{Introduction}
The critical bridge between fundamental theory and experimental data is built upon the prediction of physical observables, such as cross sections or decay rates. These observables are computed by integrating the squared matrix element ($|\mathcal{M}|^2$) over the $n$-body Lorentz-Invariant Phase Space (LIPS). As the number of final-state particles $n$ increases, the dimensionality of this integration grows rapidly, rendering traditional numerical quadrature methods intractable. Consequently, Monte Carlo (MC) techniques are the only viable tools for performing this high-dimensional integration \cite{buckley_general-purpose_2011}. These MC techniques are implemented in software frameworks known as event generators, including Herwig \cite{bellm_herwig_2016}, Sherpa \cite{gleisberg_event_2009}, Pythia \cite{sjostrand_introduction_2015}, and MadGraph5\_aMC@NLO \cite{alwall_automated_2014}, which are reviewed in \cite{buckley_monte_2019,valassi_challenges_2021}. In these generators one central task is the efficient sampling of the LIPS, as each sampled point represents a complete kinematic ``event'' of final-state particles. This kinematic sampling provides the kinematic configurations upon which the dynamics, governed by the integrand $|\mathcal{M}|^2$, are evaluated. While general-purpose algorithms capable of sampling the full LIPS exist such as RAMBO \cite{kleiss_new_1986} or GENBOD \cite{james_monte_1968}, their generality becomes a significant computational bottleneck when theoretical interest or experimental sensitivity is confined to a small sub-volume of the available phase space. This challenge arises, for example, from experimental fiducial cuts, which restrict measurements to a specific detector acceptance, as well as from dynamical effects, such as sharp resonances that concentrate the matrix element's magnitude within very small regions of phase space \cite{bothmann_efficient_2023,kleiss_weight_1994}.\\ \\
To overcome this inefficiency, standard MC techniques employ importance sampling strategies to concentrate event generation in regions where the integrand is largest. These methods, which include established adaptive multi-channel integration algorithms such as {VEGAS} \cite{lepage_adaptive_2021} and FOAM \cite{jadach_foam_2003}, as well as modern machine learning approaches such as generative neural networks (e.g., normalizing flows) \cite{butter_machine_2023,gao_event_2020,bothmann_exploring_2020}, adaptively build a sampling distribution that approximates the squared matrix element, $|\mathcal{M}|^2$. This allows for the preferential generation of events in dynamically favored regions. While highly effective, these strategies share a fundamental limitation when dealing with non-trivial kinematic constraints \cite{bothmann_exploring_2020}. 
While the physical integrand, $|\mathcal{M}|^2$, is itself Lorentz invariant, any sampling algorithm must parametrize the phase space using a specific set of coordinates to map random numbers to kinematic configurations. Event generation is almost always performed using a parametrization built from a chain of successive two-body decays (the so-called M-method), which is discussed in Section \ref{app:GENBOD}. The current samplers are thus meant to optimize the integration of the physics as a function of these two-body rest frame variables over their full and simple domains (e.g., $\cos\theta_{\mathrm{COM}} \in [-1,1]$). Experimental constraints, however, such as fiducial cuts or detector acceptance, are defined by simple geometric boundaries in the laboratory frame. Due to the non-trivial nature of the Lorentz boost, a region defined by simple, fixed-range constraints in the lab frame (e.g., $\theta_{\mathrm{lab}} \in [\theta_1, \theta_2]$) maps to a complexly bounded, non-linear, and momentum-dependent sub-volume within the two-body rest frame parametrization. 
Standard generators, which efficiently sample the entire simple domain of the two-body rest frame coordinates, have no inherent knowledge of this disjoint target sub-volume. The only recourse is to apply these laboratory frame cuts \textit{a posteriori}: one generates a large number of events across the full two-body systems domain, boosts them to the lab, and then discards the vast majority that fall outside the desired angular acceptance. This procedure is inefficient, with a computational cost that scales inversely with the acceptance volume, which can be prohibitively small. \\ \\
In this paper, we present a novel approach that directly solves the problem of inefficient laboratory frame kinematic constraints. We introduce \textsc{FANG} (Focused Angular N-body event Generator), a new Monte Carlo framework designed specifically to generate $n$-body final states \textit{a priori} within fixed, user-defined angular regions in the laboratory frame. The core of the \textsc{FANG} algorithm is a new phase space parametrization that incorporates these laboratory frame angular constraints directly into its construction. This allows the generator to sample only those kinematic configurations that satisfy the desired cuts, eliminating the need for wasteful \textit{a posteriori} rejection. Critically, this is achieved while exactly preserving the Lorentz-invariant structure of the $n$-body phase space measure, ensuring that the generated event weights are correct. This technique enables the maximally efficient computation of differential and total cross sections or decay rates for observables defined in restricted angular acceptances, offering improvements reaching up to several orders of magnitude in computational efficiency over conventional methods.\\ \\
This paper is organized as follows. Section \ref{sec:theory} reviews the LIPS formalism and the recurrence relation underlying $n$-body kinematics, and summarizes the M-generation construction that we build upon. Section \ref{sec:fang} introduces the theory and methodology of the {FANG} algorithm. Section \ref{sec:results} validates the method: (i) full–phase-space volumes against an algorithm implementing RAMBO; (ii) restricted generation against GENBOD with cuts; (iii) differential cross sections for $e\mu$ and $ep$ scattering benchmarked against analytic (Rosenbluth) results; 
and (iv) Demonstration of use with lepton–nucleus scattering models with comparison to GENIE \cite{andreopoulos_genie_2010}.
Section \ref{sec:summary} summarizes the results and outlines applications and future extensions.

\section{Theoretical Background}
\label{sec:theory}

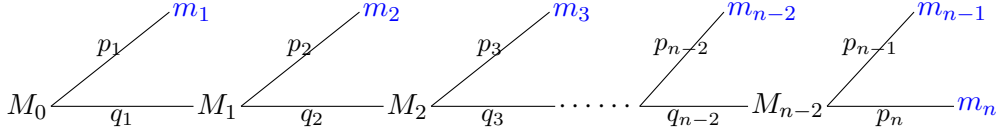
\begin{figure}[h]
\centering
\begin{tikzpicture}[
  grow=right,
  level distance=2.5cm,
  sibling distance=1.2cm,
  every node/.style={inner sep=1pt},
  edge from parent path={
    (\tikzparentnode.east) -- (\tikzchildnode.west)
  }
]

\node {$M_0$}
  child [grow=30] {node [text=blue] {$m_1$} edge from parent node[above, font=\small] {$p_1$}}
  child [grow=0] {node {$M_1$}
    child [grow=30] {node [text=blue] {$m_2$} edge from parent node[above, font=\small] {$p_2$}}
    child [grow=0] {node {$M_2$}
      child [grow=30] {node [text=blue] {$m_3$} edge from parent node[above, font=\small] {$p_3$}}
      child [grow=0] {node {$\cdots\cdots$}
          child [grow=30] {node [text=blue] {$m_{n-2}$} edge from parent node[above, font=\small] {$p_{n-2}$}}
          child [grow=0] {node {$M_{n-2}$}
            child [grow=30] {node [text=blue] {$m_{n-1}$} edge from parent node[above, font=\small] {$p_{n-1}$}}
            child [grow=0] {node [text=blue] {$m_n$} edge from parent node[below, font=\small] {$p_n$}}
            edge from parent node[below, font=\small] {$q_{n-2}$}
          }
        edge from parent node[below, font=\small] {$q_3$}
      }
      edge from parent node[below, font=\small] {$q_2$}
    }
    edge from parent node[below, font=\small] {$q_1$}
  };

\end{tikzpicture}
\caption{Generalized decay tree for \( n \) final particles,  described by $n-2$ virtual momenta \( q_i \) with virtual masses \( M_i \), and the final real on-shell momenta \( p_i \) with corresponding real masses \( m_i \).}
\label{fig:gen_tree}
\end{figure}

Both the partial decay rate of a particle of mass $M_0$ into $n$ bodies in its rest frame and the differential cross section for $2\to n$ process are given in terms of the Lorentz-invariant matrix element of the relevant process $\mathcal{M}$ and the LIPS volume element for $n$ final state particles $dV_n$,
\begin{equation}
\label{eq:decay}
d\Gamma = \frac{(2\pi)^{4-3n}}{2M_0}|\mathcal{M}|^2dV_n
\end{equation}
and 
\begin{equation}
\label{eq:dsigma}
d\sigma = \frac{(2\pi)^{4-3n}}{4\sqrt{(P_{01}\cdot P_{02})^2-M^2_{01}M^2_{02}}}|\mathcal{M}|^2dV_n
\end{equation}

where $P_{01},$ $P_{02}$, $M_{01}$ and $M_{02}$ are the corresponding four-momentum and mass of the colliding particles. The LIPS volume element for $n$ particles, $dV_n$, is given by:
\begin{equation}
\label{DLIPS}
dV_n(P; p_1, p_2, \dots, p_n) = \prod_{j=1}^n \left( d^4 p_j \, \delta(p_j^2 - m_j^2) \right) \delta^4\left( P - \sum_{j=1}^n p_j \right)\theta \left(p_j^0\right)
\end{equation}
where \( P \) is the total four-momentum of the system, which is conserved during the interaction. The four-momentum of the \( j \)-th particle in the final state is denoted by \( p_j \). The term \( \delta(p_j^2 - m_j^2) \) enforces the on-shell condition for each particle, imposing \( n \) constraints, where \( m_j \) represents the mass of the \( j \)-th particle. \( \delta^4\left( P - \sum_{j=1}^n p_j \right) \) enforces four-momentum conservation, providing 4 constraints. Together, these constraints define a \((3n - 4)\)-dimensional manifold of the \( 4n \)-dimensional space of all momentum components.
Eq. \ref{DLIPS} can be simplified to a sequential two-body decay problem, commonly referred to as the recurrence relation \cite{james_monte_1968}. Introducing invariant masses $M_i$ and corresponding 4-momentum $q_i$ which satisfy $q_i^2 = M_i^2$. We have:
{\small
 \begin{equation}
 \label{recurrence} 
dV_n(P; p_1, \dots, p_n)  = dV_2(P; p_1, q_1) \left( \prod_{i=1}^{n-3} dM_i^2 \, dV_2(q_{i}; p_{i+1}, q_{i+1}) \right) dM_{n-2}^2 \, dV_2(q_{n-2}; p_{n-1}, p_n)
 \end{equation}}
which satisfy $M_{j+1} + m_{j+1} < M_j < M_{j-1} - m_j, \forall j\in \{1, \dots, n-2 \}$ 
where we defined $M_0 =\sqrt{P^2}$. A schematic representation of the recurrence relation is shown in Fig. \ref{fig:gen_tree}.
 For the two-body phase space volume, one utilizes the property of Lorentz invariance and performs the calculation in the rest frame of the two bodies, resulting in:
\begin{equation}
\label{2body}
    dV_2(P;p_1,p_2) = \frac{1}{4M_{0}} \overset{\ast}{|\vec{p_{1}}|} d \Omega_{\overset{\ast}{p_1}}
\end{equation}
Here, the asterisk (\(* \)) denotes the two-body rest frame, where $\overset{\ast}{|\vec{p_{1}}|} = {\sqrt{\overset{\ast}{E_1^2} - m_1^2}}$ and $\overset{\ast}{E_1} = \frac{M_0^2 + m_1^2 - m_2^2}{2M_0}$ are the momentum and energy of one of the particles in the two-body rest frame, and $ d \Omega_{\overset{\ast}{p_1}}$  is the solid angle element of $\overset{\ast}{p_1}$ in the two-body rest frame. 
Note that Eq. \ref{2body} has 2 degrees of freedom, the azimuthal and polar coordinates, which aligns with the analysis made earlier, indicating that the system should have \( 3n - 4 \) degrees of freedom. This is a special case for \( n = 2 \).
Alternatively, Eq. \ref{2body} may be written directly as a function of the particle masses \cite{kleiss_monte_2019}:
\begin{equation}
    dV_2(P;p_1,p_2) = \frac{1}{8} F \left( \frac{m_1^2}{M_0^2}, \frac{m_2^2}{M_0^2} \right)  d \Omega_{\overset{\ast}{p_1}} 
\end{equation}
similarly  the 2 body decay of the rest frame of particle $q_i$ to 2 bodies can be written as
\begin{align} 
   dV_2(q_{i}; p_{i+1}, q_{i+1})   &= \frac{1}{8} F \left( \frac{m_{i+1}^2}{M_{i}^2}, \frac{M_{i+1}^2}{M_i^2} \right)  d \Omega_{\overset{\ast}{p}_{i+1}}
\end{align}   

Where $F(x, y) = \left( (1 - x - y)^2 - 4xy \right)^{1/2}$.

Inserting the two-body phase space elements back into Eq. \ref{recurrence} yields,
{\footnotesize

\begin{align}
dV_n =
\left(\frac{1}{2}\right)^{3n-3} F \left( \frac{m_1^2}{M_0^2}, \frac{M_1^2}{M_0^2} \right)   \left( \prod_{i=1}^{n-3}dM_i^2 \,  F \left( \frac{m_{i+1}^2}{M_i^2}, \frac{M_{i+1}^2}{M_i^2} \right) d\Omega_{\overset{\ast}{p}_{i+1}}  \right)dM_{n-2}^2 F \left( \frac{m_{n-1}^2}{M_{n-2}^2}, \frac{m_n^2}{M_{n-2}^2} \right) d\Omega_{\overset{\ast}{p_1}}  d\Omega_{\overset{\ast}{p}_{n-1}} 
\label{eq:dOmega}
\end{align}}

In case of full phase space, integrating over the solid angle in each two-body rest frame, provides a factor $(4\pi)^{n-1}$  and the phase space element is reduced to:
{\footnotesize
\begin{align}
    &dV_n^{4\pi} =
    \frac{\pi^{n-1}}{2} F \left( \frac{m_1^2}{M_0^2}, \frac{M_1^2}{M_0^2} \right)   \left( \prod_{i=1}^{n-3} dM_i \,  M_i  F \left( \frac{m_{i+1}^2}{M_i^2}, \frac{M_{i+1}^2}{M_i^2} \right)   \right)dM_{n-2} \, M_{n-2} F \left( \frac{m_{n-1}^2}{M_{n-2}^2}, \frac{m_n^2}{M_{n-2}^2} \right)
    \label{eq:nbodyPS}
\end{align}}

\subsection{M-generation overview}
\label{app:GENBOD}
FANG is based on the $M$-generation for $n$-body phase space generation, originally implemented in the code GENBOD \cite{james_monte_1968}. The  GENBOD technique appears to have been independently developed by Kopylov \cite{kopylov_notitle_1960} in the Soviet Union and by Raubold and Lynch at CERN.
The $M$-generation algorithm treats the decay as a sequence of two-body decays (see Fig.~\ref{fig:gen_tree}), starting from the system mass $M_{0}$ and ending with the final two outgoing particles. For $n$ outgoing particles of masses $m_{i}$, there are $n-2$ intermediate virtual particles of masses $M_{i}$ that must be generated. The generation of the virtual masses $M_{i}$ proceeds according to the following prescription:
\begin{itemize}
    \item Generate $n-2$ random numbers $r_{i}$ uniformly distributed between 0 and 1.
    \item Order these numbers in decreasing magnitude, $r_{i} > r_{i+1}$.
    \item Compute the virtual masses according to
\end{itemize}

\begin{equation}
    M_{i} = r_{i}\!\left(M_{0} - \sum_{j=1}^{i} m_{j}\right) + \sum_{j=i+1}^{n} m_{j}.
\end{equation}

This procedure provides an efficient means of generating the virtual masses $M_{i}$, ensuring that they are chosen independently and always lie within the physical region, i.e.,
\begin{equation}
    M_{j-1} - m_{j} \; > \; M_{j} \; > \; M_{j+1} + m_{j+1}.
\end{equation}

Starting from the $M_{0}$ rest frame, each two-body decay is treated by sampling $\cos\theta$ uniformly in the range $[-1,1]$ and $\varphi$ uniformly in the range $[0,2\pi]$ in the two-body rest frame, ensuring an isotropic phase space. The resulting four-momenta are then boosted back to the laboratory frame, after which the subsequent two-body decay can be evaluated.
Analytically, the phase space for an $n$-body decay is obtained by integrating the differential volume element in Eq.~\ref{eq:nbodyPS} over the masses within the physical region. In the Monte Carlo approach, each event is assigned a weight equal to the integrand of Eq.~\ref{eq:nbodyPS}. The phase space is then estimated by summing all weights corresponding to events in the physical region, multiplying by the volume spanned by the generated random numbers, and dividing by the total number of generated events. The range $\Delta M$ over which each virtual mass is generated is the same for all masses and is given by 
\[
\Delta M = M_{0} - \sum_{i=1}^{n} m_{i},
\] 
so that the total volume is $\Delta M^{\,n-2}$.

However, since the recipe employs the ``trick'' of ordering the random numbers, which guarantees that all generated masses lie within the physical region, it is necessary to include a combinatorial correction factor of $(n-2)!$. Thus, the weight $W_{i}$ assigned to each event is

\begin{equation}
\label{eq:Wi4pi}
     W_i = 
    \frac{\Delta M^{n-2}}{(n-2)!}\frac{\pi^{n-1}}{2} F \left( \frac{m_1^2}{M_0^2}, \frac{M_1^2}{M_0^2} \right)   \left( \prod_{i=1}^{n-3} \,  M_i  F \left( \frac{m_{i+1}^2}{M_i^2}, \frac{M_{i+1}^2}{M_i^2} \right)   \right)\, M_{n-2} F \left( \frac{m_{n-1}^2}{M_{n-2}^2}, \frac{m_n^2}{M_{n-2}^2} \right)
\end{equation}
For a total of $N$ Monte Carlo events, the $n$-body phase space volume $V_{n}$ and its corresponding statistical uncertainty are evaluated as
\begin{equation}
 V_{n}=\frac{1}{N}\sum_{i=1}^{N}W_i \pm \frac{1}{N}\sqrt{\sum_{i=1}^{N}W_{i}^{2}}
 \label{eq:Vn}
\end{equation}

If the matrix element $\mathcal{M}$ is known and can be evaluated from the four-momenta of the generated particles in each event, the decay rate and cross section can be computed using Eqs.~\ref{eq:decay} and \ref{eq:dsigma}, respectively, as
\begin{equation}
\label{eq:fdecay}
\Gamma = \frac{(2\pi)^{4-3n}}{2M_0}\frac{1}{N}\sum_{i=1}^{N}W_i|\mathcal{M}_i|^2
\end{equation}
and 
\begin{equation}
\label{eq:fsigma}
\sigma = \frac{(2\pi)^{4-3n}}{4\sqrt{(P_{01}\cdot P_{02})^2-M^2_{01 }M^2_{02}}} \frac{1}{N}\sum_{i=1}^{N}W_i|\mathcal{M}_i|^2
\end{equation}

\section{FANG Approach}
\label{sec:fang}
\subsection{Differential cross section in the laboratory frame}

To calculate the differential angular cross section of $n_d<n$ particles in fixed (constrained) $n_d$ directions in the laboratory frame, it is necessary to replace each factor of $d\Omega_{\overset{\ast}{p_k}}$ in Eq.~\ref{eq:dOmega}
 by
\begin{equation}
d\Omega_{\overset{\ast}{p_k}} = J_k \, d\Omega_{p_k}, \qquad k \in \{1,2,\dots,n_d\},
\end{equation}

where $J_k$ is the Jacobian relating the solid angle in the rest frame of the two bodies to that in the laboratory frame.
\begin{equation}
\label{eq:jac}
J_k \;=\; \left| \frac{d \Omega_{\overset{\ast}{p_k}}}{d\Omega_{p_k}} \right| 
\;=\; \left| \det \begin{pmatrix} 
\dfrac{\partial \cos{\overset{\ast}{\theta}_k}}{\partial \cos{\theta_k}} & 
\dfrac{\partial \cos{\overset{\ast}{\theta}_k}}{\partial \varphi_k} \\[1.2ex] 
\dfrac{\partial \overset{\ast}{\varphi}_k}{\partial \cos{\theta_k}} & 
\dfrac{\partial \overset{\ast}{\varphi}_k}{\partial \varphi_k} 
\end{pmatrix} \right|
\end{equation}
with $\theta_k$ and $\varphi_k$ denoting the polar and azimuthal angles of the momentum of particle $k$ in the laboratory frame, and $\overset{\ast}{\theta}_k$ and $\overset{\ast}{\varphi}_k$ the corresponding momentum angles in the two-body rest frame, all angles defined relative to this two-body system momentum direction. 
For each fixed (constrained) emitted laboratory directions $(\hat{r}_1, \hat{r}_2, \dots, \hat{r}_{n_d})$,
the Jacobian is evaluated at the angular coordinates $(\theta_k,\varphi_k)$ corresponding to the laboratory direction $\hat{r}_k$. For this pair of laboratory angles, the quantities $\overset{\ast}{\theta}_k$ and $\overset{\ast}{\varphi}_k$ are then determined according to the procedure described in Section~\ref{sec:4momConst}.
 \\
 since the azimuthal angle is unaffected by the boost, Eq.~\ref{eq:jac} is reduced to
\begin{equation}
J_k  = \left| \frac{\partial \cos{\overset{\ast}{\theta}_{k}}}{\partial \cos{\theta_{k}}} \right|
\end{equation}
Computing this derivative yields \cite{horin_development_2025}:
\begin{equation}
 J_k =\left| \frac{\left(\frac{\sin{\overset{\ast}{\theta}_{k} }}{\sin{{\theta}_{k} }}\right)^{3}}
{\gamma\left(1+\frac{\beta}{\overset{\ast}{\beta_{k}}}  \cos {\overset{\ast}{\theta}_{k} }\right)  } \right|
\label{eq:J}
\end{equation}
This expression for $J_k$ fails numerically at $\theta_{k} \to 0$, but the following limit is obtained \cite{braithwaite_relativistic_1972}:
\begin{equation}
    {J_k} =\left[\frac{\gamma \cos {\overset{\ast}{\theta}}_{k}\left[1+\left(\frac{\beta}{\overset{*}{\beta}_{k}}\right) \cos {\overset{\ast}{\theta_{k}}}\right]}{\cos {\theta_{k}}}\right]^2 \text { for } \theta_{k} \to 0
\end{equation}

The $\gamma$ is the Lorentz factor of the two-body system in the laboratory frame,  \( \overset{*}{\beta}_k \) is the velocity of the 
$k^{th}$ constrained ejected particle in the two-body rest frame, and $\beta$ is the velocity of the two-body system in the laboratory frame.

Returning to  Eq. \ref{eq:dOmega}, integrating this time only over solid angles of the unconstrained ejected particles and expressing the solid angle in the laboratory frame, the phase space element is expressed as
\begin{align}
    \label{eq:nbodyCPS}
dV_n^{\Delta \Omega}  =   &\frac{\pi^{n-1}}{2} F \left( \frac{m_1^2}{M_0^2}, \frac{M_1^2}{M_0^2} \right)   \left( \prod_{i=1}^{n-3} dM_i \,  M_i  F \left( \frac{m_{i+1}^2}{M_i^2}, \frac{M_{i+1}^2}{M_i^2} \right)   \right)dM_{n-2} & \\
 &M_{n-2} F \left( \frac{m_{n-1}^2}{M_{n-2}^2}, \frac{m_n^2}{M_{n-2}^2} \right)\prod_{k=1}^{n_d} \left(\frac{J_k}{4\pi} d\Omega_k \right) & \nonumber
\end{align}

inserting $dV_n^{\Delta \Omega}$ in Eq. \ref{eq:dsigma} and integrating over the virtual masses $M_i$, provides the differential cross section
\begin{align}
\label{eq:dsdO}
\frac{d^{n_d}\sigma}{d\Omega_1 \dots d\Omega_{n_d}} = &\frac{(2\pi)^{4-3n}}{4\sqrt{(P_{01}\cdot P_{02})^2-M^2_{01}M^2_{02}}}  \frac{\pi^{n-1}}{2} \int_{M_{1}\dots M_{n-2}} |\mathcal{M}|^2F \left( \frac{m_1^2}{M_0^2}, \frac{M_1^2}{M_0^2} \right)&  \\
 & \times   \left( \prod_{i=1}^{n-3} dM_i \,  M_i  F \left( \frac{m_{i+1}^2}{M_i^2}, \frac{M_{i+1}^2}{M_i^2} \right)   \right)  dM_{n-2}M_{n-2} F \left( \frac{m_{n-1}^2}{M_{n-2}^2}, \frac{m_n^2}{M_{n-2}^2} \right)\prod_{k=1}^{n_d} \left(\frac{J_k}{4\pi} \right) &\nonumber
\end{align}

 \subsection{Calculating the four-momentum of the constrained particle}
 \label{sec:4momConst}

Consider the first vertex in the decay chain shown in Fig.~\ref{fig:gen_tree}, where a system of invariant mass $M_{0}$ and four-momentum $P$ decays into a virtual system of mass $M_{1}$ and an outgoing particle of mass $m_{1}$ with four-momentum $p_{1}$. When $p_{1}$ is constrained to lie along the laboratory direction $\hat{r}_{1}$, the magnitude of the three-momentum of the outgoing particle in the $M_{0}$ rest frame, $\overset{\ast}{|\vec{p}_{1}|}$, and in the laboratory frame, $|\vec{p}_{1}|$, are given by

\begin{equation}
\label{eqn:p3_2form}
    \begin{split}
\overset{\ast}{|\vec{p_1}|} &=\frac{\sqrt{Z^2-M_{0}^2m_1^2}}{M_{0}} \\
|\vec{p_1}|  &= \frac{ (|\vec{P}|Z\cos{\theta}) \pm E_{0}\sqrt{|\overset{\ast}{\vec{p_1}}|^{2} M_0^2- m_{1}^2\sin^{2}{\theta}|\vec{P}|^{2} } } {(|\vec{P}|^{2}\sin^{2}{\theta}+M_{0}^2)}
\end{split}
\end{equation}
where we denote the initial energy as $E_{0} = \sqrt{M_{0}^{2} + |\vec{P}|^{2}}$, define $Z = \tfrac{M_{0}^{2} + m_{1}^{2} - M_{1}^{2}}{2}$, and let $\theta$ be the angle between $\vec{P}$ and $\hat{r}_{1}$ in the laboratory frame.\\
 The momentum of $m_{1}$ in the laboratory frame, $|\vec{p}_{1}|\hat{r}_{1}$, can be boosted to the $M_{0}$ rest frame and used to calculate the momentum of $M_{1}$, $\overset{\ast}{\vec{q}}_{1}$, which is then boosted back to the laboratory frame to obtain $\vec{q}_{1}$.
Once the momentum of $M_{1}$, $\vec{q}_{1}$, is determined in the laboratory frame, the kinematics of the subsequent vertex can be calculated, and the procedure continued accordingly.

The momentum $|\vec{p}_{1}|$ may admit zero, one, or two physical solutions. A solution is considered physical if $|\vec{p}_{1}|$ is real and positive. Events with no physical solution are discarded, while in cases with multiple solutions all physical ones are retained.
For example, in the case of $n=3$, if two particles are constrained and the first vertex admits two possible solutions, and for each of these the second vertex also yields two solutions, then four distinct events are generated corresponding to the same set of virtual masses $M_{i}$.

\subsection{Methodology}
Building on the $M$-generation described in Section \ref{app:GENBOD}, we extend the algorithm to allow, within an $n$-body phase space, laboratory-frame directional constraints on $n_{d}<n$ particles. This extension enables the generation of only those events consistent with the imposed constraints, while preserving exact kinematics and providing an accurate evaluation of the relative phase-space volume.
For differential calculations, the constrained particle is generated in all events along the same laboratory-frame direction, specified by the vector $\hat{r}_{k}$. The kinematics of all other particles in the decay are, of course, influenced by this constraint.
For partial phase-space event generation, where the constrained particle is restricted to a finite solid angle $\Delta\Omega_{k}$ in the laboratory frame, a new direction vector $\hat{r}_{k}$ is generated for each event, isotropically distributed within the limits of $\Delta\Omega_{k}$. This vector $\hat{r}_{k}$ is then used in the calculation of the four-momentum of the constrained particle.
The inputs to the algorithm are the initial system four-momentum $P$, the outgoing particle masses $m_{i}$, and, for each constrained particle, the direction $\hat{r}_{k}$ together with the size and shape of the constraining solid angle $\Delta\Omega_{k}$. The algorithm then proceeds according to the following steps:

\begin{itemize}
    \item Arrange the outgoing particles such that the constrained particles appear first.
    \item Generate the virtual masses $M_{i}$ according to the standard $M$-generation procedure.
    \item Determine the four-momentum of the constrained particle; there may be 0, 1, or 2 physical solutions:
    \begin{itemize}
        \item If no physical solution exists, reject the event.
        \item If more than one solution exists, generate all corresponding events.
    \end{itemize}
    \item Compute the Jacobian for the constrained particles (see Eq.~\ref{eq:J}).
    \item Generate the remaining unconstrained particles using the standard $M$-generation procedure.
    \item Evaluate the weight of the event.
\end{itemize}

\subsubsection{Event weight calculation}
After determining the four-momenta of the constrained particles as described above, the corresponding Jacobians can be evaluated using Eq.~\ref{eq:J}. With the Jacobians known, the event weight can then be computed.
For $n_{d}$ constrained particles, based on Eq.~\ref{eq:nbodyCPS} and including the combinatorial correction factor $(n-2)!$, each event is assigned the weight

\begin{equation} 
\label{eq:Wi}
     W_i = 
    \frac{\Delta M^{n-2}}{(n-2)!}\frac{\pi^{n-1}}{2} F \left( \frac{m_1^2}{M_0^2}, \frac{M_1^2}{M_0^2} \right)   \left( \prod_{i=1}^{n-3} \,  M_i  F \left( \frac{m_{i+1}^2}{M_i^2}, \frac{M_{i+1}^2}{M_i^2} \right)   \right)\, M_{n-2} F \left( \frac{m_{n-1}^2}{M_{n-2}^2}, \frac{m_n^2}{M_{n-2}^2} \right)\prod_{k=1}^{n_d} \left(\frac{J_k}{4\pi} \right)
\end{equation}
Equation \ref{eq:Wi4pi} is recovered as a special case of Eq. \ref{eq:Wi} when $n_d=0$.

\subsubsection{Integration over Detector geometry}
\label{app:strips}
For partial phase-space event generation, where a constrained particle is restricted to a finite solid angle $\Delta\Omega_{k}$ in the laboratory frame, a new constraining vector $\hat{r}_{k}$ is generated isotopically  in the laboratory frame within $\Delta\Omega_{k}$ for each event. 
In these cases, the event phase space restricted to finite solid angles is given by the angle-integrated form of Eq. \ref{eq:Vn}.  
\begin{equation}
 V^{\Delta\Omega}_{n} = \prod_{k=1}^{n_{d}} \Delta\Omega_{k} \, \frac{1}{N} \sum_{i=1}^{N} W_{i},
 \label{eq:VnO}
\end{equation}
Where $W_i$ is the weight of each event  given by Eq. \ref{eq:Wi}.

The algorithm permits generation of the constrained particles in five different configurations.

For differential calculations, the algorithm supports the following routines:  
\begin{itemize}
    \item \textbf{Point generation} - the particle is always generated in the same specified laboratory-frame direction $\hat{r}_{k}$.
    \item \textbf{Ring generation} - in the laboratory frame, the particle is generated with a fixed polar angle $\theta_{k}$ relative to a specified direction $\hat{r}_{k}$, while the azimuthal angle $\varphi'$ around $\hat{r}_{k}$ is sampled uniformly in the range $[0,2\pi]$.
\end{itemize}

For partial phase-space event generation and calculation, the following configurations are supported:  
\begin{itemize}
    \item \textbf{Strip generation} - in the laboratory frame, the particle is generated within a solid angle defined by two independent random numbers: $\cos\theta$ in the range $\cos\theta_{0} - \tfrac{\Delta \cos}{2} < \cos\theta < \cos\theta_{0} + \tfrac{\Delta \cos}{2}$ and $\varphi$ in the range $\varphi_{0} - \tfrac{\Delta \varphi}{2} < \varphi < \varphi_{0} + \tfrac{\Delta \varphi}{2}$, with $\Delta\Omega = \Delta \cos \, \Delta \varphi$.  

    \item \textbf{Circle generation} - in the laboratory frame, the particle is generated within a cone around a specified central direction. Two random numbers are used: $\cos\theta'$ in the range $1 - \Delta \cos < \cos\theta' < 1$, where $\theta'$ is the polar angle relative to the central direction of the solid angle, and $\varphi'$ in the range $0 \leq \varphi' < 2\pi$, which is the azimuthal angle around that direction. The solid angle is given by $\Delta\Omega = 2\pi \, \Delta \cos$.  

    \item \textbf{User-defined solid angle} - this option generalizes point generation. Instead of fixing the same direction for each event, the user supplies a different constraining vector $\hat{r}_{k}$ for each event, generated isotropically within a specified solid angle in the laboratory frame. Alternatively, the user may choose to generate the constraining vectors anisotropically (e.g., following the $|\mathcal{M}|^{2}$ dependence) and apply weights to restore isotropy.  
\end{itemize}

\subsubsection{Differential calculation}
The differential phase space in the specified laboratory-frame directions $\hat{r}_{k}$ is evaluated using point generation, where the constrained $k^{\text{th}}$ particle is directed along the same $\hat{r}_{k}$ for all events. The resulting differential phase space is given by

\begin{equation}
\frac{d^{n_d}V_n(\hat{r}_1, \dots,\hat{r}_{n_d})}{d\Omega_1, \dots,d\Omega_{n_d}} =  \frac{1}{N}\sum_{i=1}^{N}W_i
\end{equation}
For a known matrix element $\mathcal{M}$, the differential cross section in the specified laboratory-frame directions $\hat{r}_{k}$ can be calculated as
\begin{equation}
\label{eq:dsigmadO}
\frac{d^{n_d}\sigma(\hat{r}_1, \dots,\hat{r}_{n_d})}{d\Omega_1, \dots,d\Omega_{n_d}} = \frac{(2\pi)^{4-3n}}{4\sqrt{(P_{01}\cdot P_{02})^2-M^2_{01 }M^2_{02}}} \frac{1}{N}\sum_{i=1}^{N}W_i|\mathcal{M}_i|^2
\end{equation}

It is also possible to employ ring generation, whereby the azimuthal angle $\varphi$ is integrated over to extract $\tfrac{dV_{n}}{d\cos\theta}$,

\begin{equation}
\label{eq:dVdcos}
\frac{d^{n_d}V_n(\theta_1, \dots,\theta_{n_d})}{d\cos{\theta_1}, \dots,d\cos{\theta_{n_d}}} = (2\pi)^{n_d}\frac{1}{N}\sum_{i=1}^{N}W_i
\end{equation}

and similarly, for a known matrix element $\mathcal{M}$, the differential cross section is

\begin{equation}
\label{eq:dsigmadcos}
\frac{d^{n_d}\sigma(\theta_1, \dots,\theta_{n_d})}{d\cos{\theta_1}, \dots,d\cos{\theta_{n_d}}} = \frac{(2\pi)^{4-3n}(2\pi)^{n_d}}{4\sqrt{(P_{01}\cdot P_{02})^2-M^2_{01 }M^2_{02}}} \frac{1}{N}\sum_{i=1}^{N}W_i|\mathcal{M}_i|^2\, .
\end{equation}

This method provides an efficient and accurate means of calculating differential cross sections from matrix elements, applicable to any number of final-state particles and for arbitrary directional constraints.

\section{Results}
\label{sec:results}
\subsection{Full phase space calculation}

Testing the full phase-space calculation verifies that the $M$-generation is correctly implemented in the FANG algorithm. This is done by comparing full phase-space volumes computed with FANG (see Eq.~\ref{eq:Vn}) to those obtained with HAZMA ~\cite{coogan_hazma_2020}, which is an implementation of the widely used event generator RAMBO~\cite{kleiss_new_1986}.
The results summarized in Table~\ref{tab:tests} show excellent agreement between the two codes. In most cases, RAMBO is slightly more efficient than FANG, although at the non-relativistic limit, FANG becomes comparatively more competitive.

\begin{table}[H]
\centering
\caption{Phase-space volume calculations for $P(P_{x},P_{y},P_{z},E)=(0,0,5,13)$ (same units as mass),   decaying into final states with the specified masses. The HAZMA results have been multiplied by $(2\pi)^{3n-4}$.}

\label{tab:tests}
\begin{tabular}{|c|c|c|c|}
\hline
Number  &Masses  & FANG   &HAZMA  \\
of bodies& values & $10^8$ events    &$10^8$ events \\
\hline 
3 &{1,1,1}&$141.469 \pm 0.015$  & $141.4581 \pm 0.0020$ \\ \hline
4 &{1,1,1,1}& $3285.76 \pm 0.36$ & $3286.81 \pm 0.10$ \\ \hline
5 &{1,1,1,1,1}& $26628.1 \pm 3.0$&  $26630.0 \pm 1.2$ \\ \hline
6 & {1,1,1,1,1,1}& $80467.7 \pm 9.4$&  $80499.3 \pm 5.0$ \\
 \hline
6 &{2,2,2,2,2,1.9}& $ 4.87011\times 10^{-7} \pm 5.6\times 10^{-11} $ & $4.87115 \times 10^{-7} \pm 5.8 \times 10^{-11}$ \\
 \hline
\end{tabular}
\end{table}

\subsection{Partial phase space event generation}
To test the partial phase-space event generation, the angular and energy distributions of the outgoing particles are compared between FANG, which generates only events within the designated solid angles, and a full phase-space Monte Carlo, where cuts are applied to select events in which the restricted particles fall within the designated solid angles.
The full phase-space Monte Carlo used for this comparison is GENBOD, as implemented in the \texttt{TGenPhaseSpace} class of the ROOT Data Analysis Framework~\cite{canal_root-projectroot_2025}. The test scenario considers a parent particle with four-momentum $P(P_{x},P_{y},P_{z},E)=(0,0,5,13)$ decaying into five particles, each of mass $m_{i}=1$, at the origin of the laboratory coordinate system.
 Three of the particles are constrained. The first, with momentum $p_{1}$, is required to reach a circular ``detector'' of radius 0.2 centered along the $z$ axis at the coordinates $(0,0,0.5)$ (same units of length as the radius). The second, with momentum $p_{2}$, is required to reach a circular ``detector'' of radius 0.3 centered along the $x$ axis at the coordinates $(0.5,0,0)$. The third particle is constrained to lie within a solid-angle strip defined by $-0.75 < \cos\theta < 0.75$ and $0.1\pi < \varphi < 0.9\pi$.
 The phase space of the unconstrained full Monte Carlo is given in Table~\ref{tab:tests} as $26628.1 \pm 3.0$, while the phase space of the constrained Monte Carlo, calculated using Eq.~\ref{eq:VnO}, is $4.4544 \pm 0.0061$, i.e., about $\sim 5000$ times smaller. This implies that achieving the same statistical precision would require running the full Monte Carlo with roughly $5000$ times more events than FANG. For this comparison, a relatively large solid angle was deliberately chosen; however, in some scenarios the constrained phase space can be as small as $10^{-9}$ of the full phase space, in which case relying on a full Monte Carlo with cuts becomes impractical.

\begin{figure}[H]
    \centering
    \includegraphics[width=0.8\linewidth]{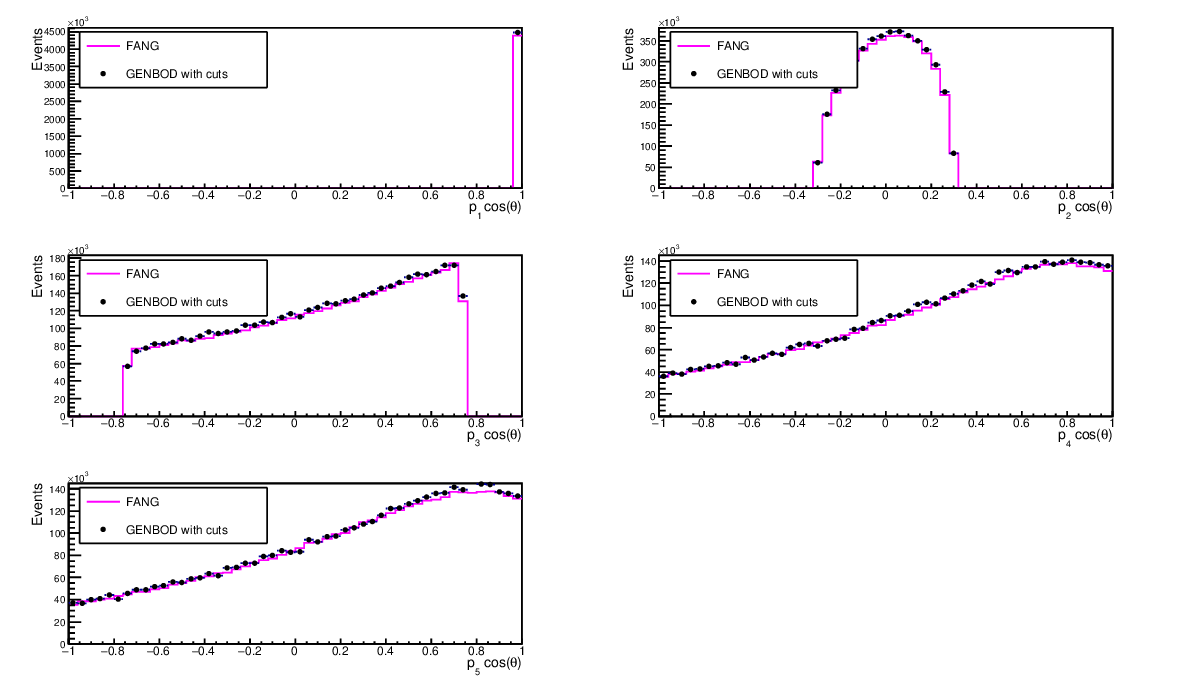}
   \caption{Comparison of the $\cos\theta$ distributions for all five particles between the full simulation with cuts and the FANG algorithm, which generates only events within the designated solid angles.}
    \label{fig:compCos}
\end{figure}

\begin{figure}[H]
    \centering
    \includegraphics[width=0.8\linewidth]{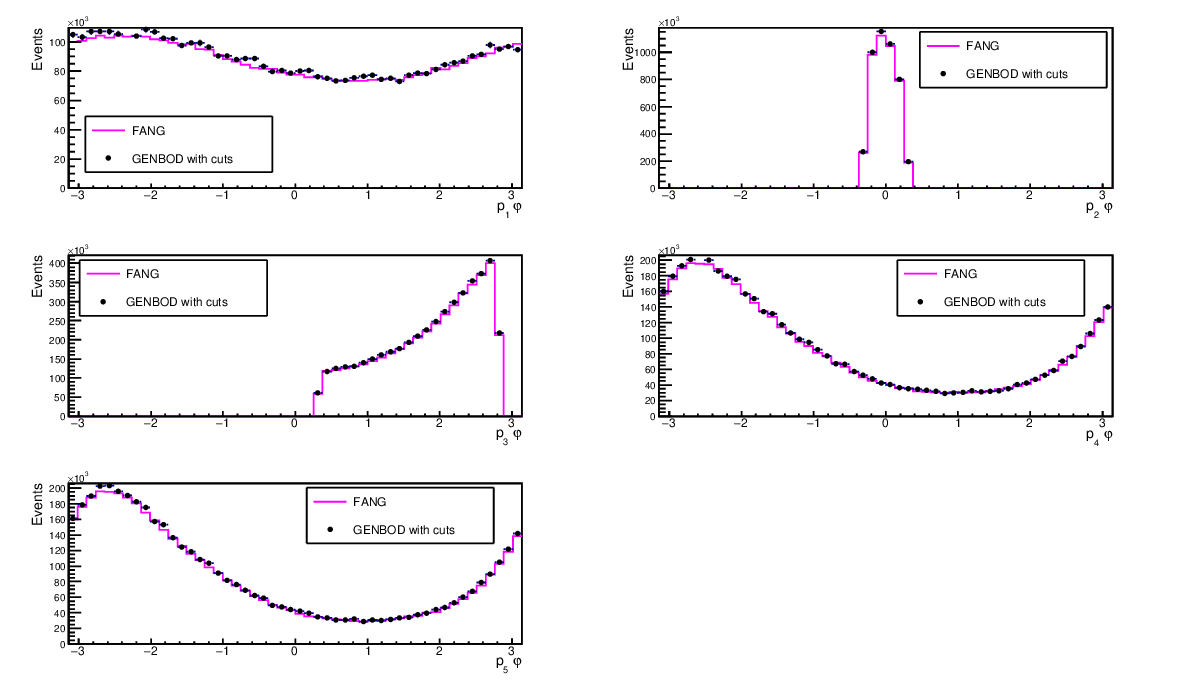}
   \caption{Comparison of the $\varphi$ distributions for all five particles between the full simulation with cuts and the FANG algorithm, which generates only events within the designated solid angles.}
    \label{fig:compPhi}
\end{figure}

\begin{figure}[H]
    \centering
    \includegraphics[width=1\linewidth]{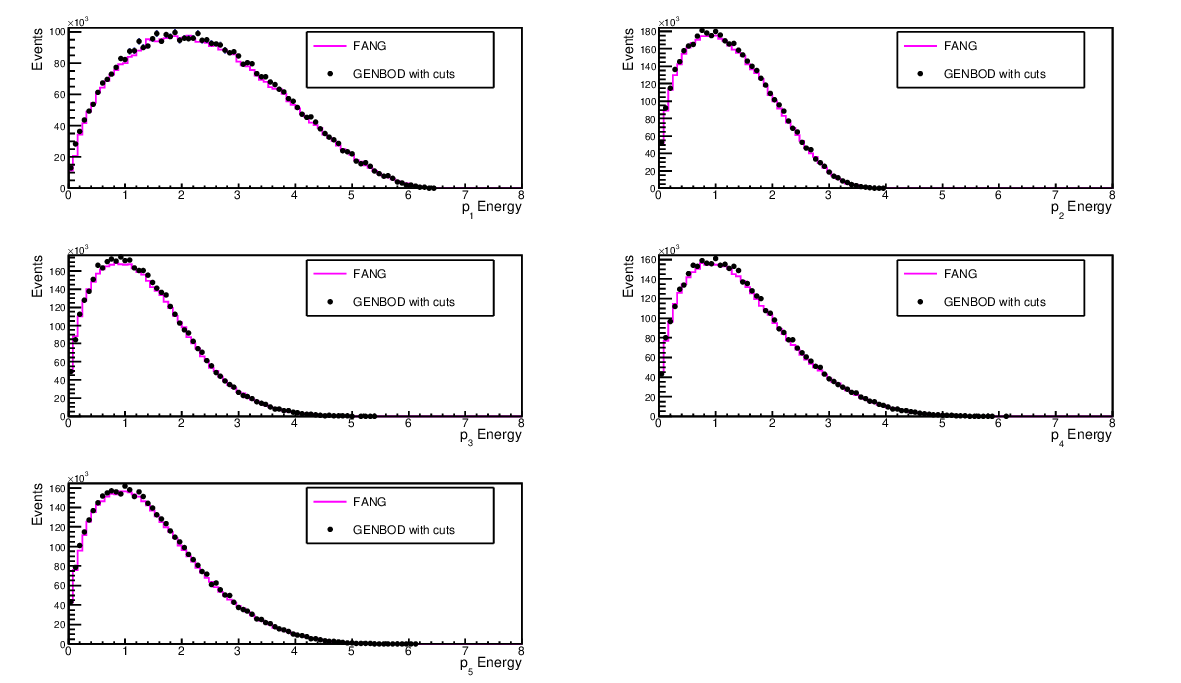}
\caption{Comparison of the energy distributions (same units as mass) for all five particles between the full simulation with cuts and the FANG algorithm, which generates only events within the designated solid angles.}
    \label{fig:compE}
\end{figure}

Fig.~\ref{fig:compCos} shows perfect agreement between the $\cos\theta$ distributions of all five particles obtained from the full simulation with cuts and from the FANG algorithm, which generates only events within the designated solid angles.
For the tests presented here, FANG was run with $10^{6}$ events, while the full GENBOD simulation required $10^{9}$ events and still yielded five times fewer statistics after cuts. Similar levels of agreement are observed in the $\varphi$ distributions (Fig.~\ref{fig:compPhi}) and in the energy distributions (Fig.~\ref{fig:compE}).

\subsection{Differential cross section calculation}
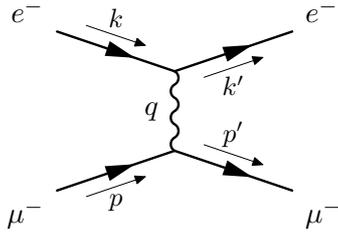
\begin{figure}[H]
    \centering

\begin{fmffile}{feyngraph}
  \begin{fmfgraph*}(110,60)
    \fmfleft{i1,i2}
    \fmfright{o1,o2}
    \fmflabel{$e^-$}{i2}
    \fmflabel{$e^-$}{o2}
    \fmflabel{$\mu^-$}{i1}
    \fmflabel{$\mu^-$}{o1}
    \fmf{fermion}{i2,v2,o2}
    \marrow{ea}{ up }{top}{$k$}{i2,v2}
    \marrow{eb}{down}{bot}{$k'$}{v2,o2}
    \fmf{fermion}{i1,v1,o1}
    \marrow{ma}{down}{bot}{$p$}{i1,v1}
    \marrow{mb}{ up }{top}{$p'$}{v1,o1}
    \fmf{photon,label=$q$}{v2,v1}
  \end{fmfgraph*}
\end{fmffile}
\caption{Feynman diagram for $e^{-}\mu^{-} \rightarrow e^{-}\mu^{-}$ scattering.}

\label{fig:fynman}
\end{figure}

Having established the correctness of the total and partial phase-space integration, the next step is to verify that the method also reproduces physical observables by integrating the phase space with the squared matrix amplitude, as required for the calculation of differential cross sections.
The differential cross section calculation is tested using a textbook example of $e^{-}\mu^{-} \rightarrow e^{-}\mu^{-}$ scattering taken from \textit{Quarks \& Leptons} by Halzen and Martin~\cite{halzen_quark_2008} (see Fig.~\ref{fig:fynman}). The differential cross section for this process can be derived analytically and is therefore well suited for comparison with calculations performed using the FANG algorithm.

The spin-averaged $e^-\mu^-\longrightarrow e^-\mu^-$ squared amplitude for this process, as given in Ref.~\cite{halzen_quark_2008}, is calculated to be 
\begin{equation}
\label{eq:emuAmp}
\overline{|\mathcal{M}|^2}_{e \mu \to e \mu}=\frac{128\pi^2\alpha^2}{q^4} \left[ (k'\cdot p')(k\cdot p) + (k'\cdot p)(k\cdot p') -m_e^2p' \cdot p -m_{\mu}^2 k' \cdot k + 2 m_e^2 m_{\mu}^2  \right]
\end{equation}
where $p$ and $p'$ are the muon four-momenta before and after scattering, respectively,  
$k$ and $k'$ are the corresponding electron four-momenta,  
$q = k' - k$ is the momentum transfer,  
and $m_{e}$ and $m_{\mu}$ are the electron and muon masses.
Neglecting all terms proportional to the electron mass $m_{e}$, the differential elastic cross section for $e^{-}\mu^{-}$ scattering in the laboratory frame is obtained as
\begin{equation}
\label{eq:formula}
\frac{d\sigma}{d\Omega}_{lab} = \left( \frac{\alpha^2}{4E^2\sin^4{\frac{\theta}{2}}} \right) \frac{E'}{E}\left( \cos^2{\frac{\theta}{2}}-\frac{q^2}{2m_{\mu}^2}\sin^2{\frac{\theta}{2}}  \right)
\end{equation}
Here $E$ and $E'$ denote the energies of the incoming and outgoing electrons, respectively, and the laboratory frame is defined as the frame in which the muon is initially at rest.

\begin{figure}[H]
    \centering
    \includegraphics[width=0.8\linewidth]{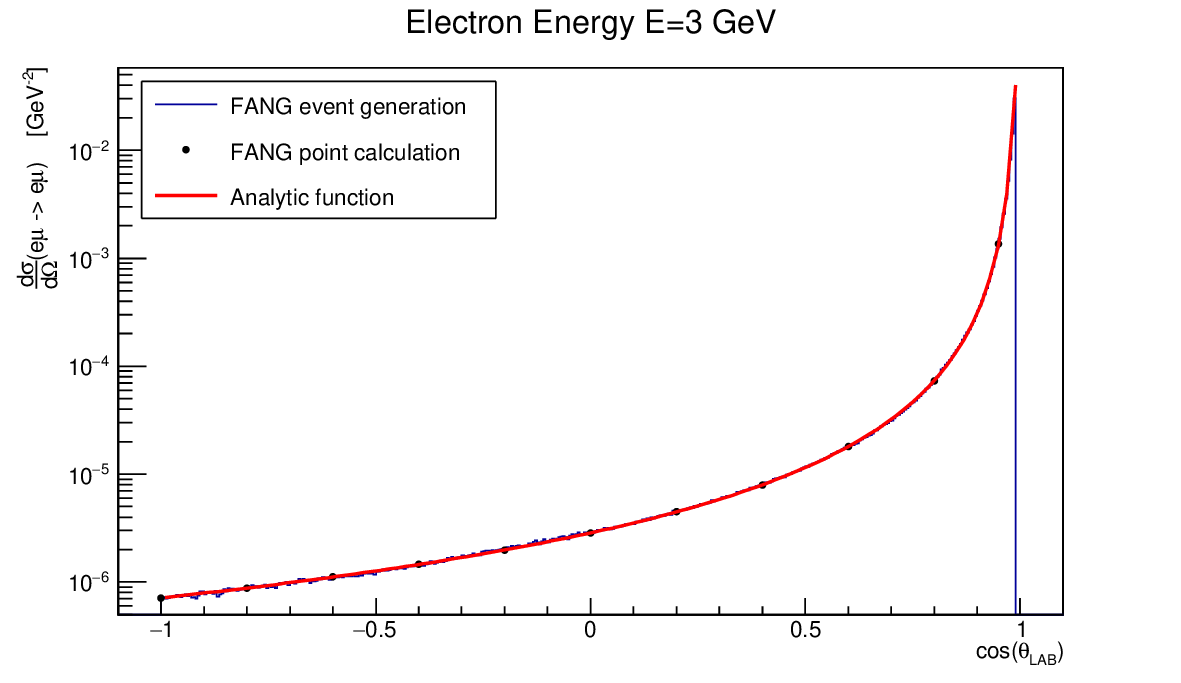}
    \caption{Differential $e^{-}\mu^{-}$ elastic cross section, showing agreement between the analytic expression of Eq.~\ref{eq:formula} (red line), the FANG calculation (black points) and the (blue histogram) of events generated by FANG.}

    \label{fig:dSdO-3GeV}
\end{figure}

\begin{figure}[H]
    \centering
    \includegraphics[width=0.8\linewidth]{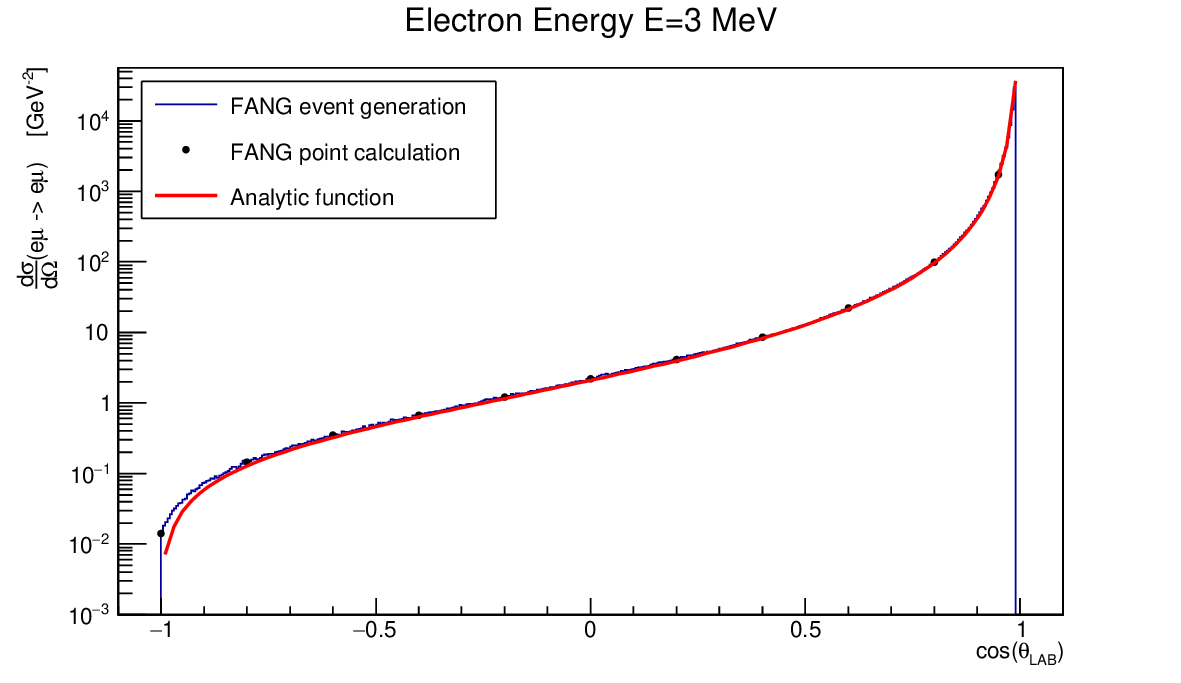}
    \caption{Expected divergence in the differential $e^{-}\mu^{-}$ elastic cross section between the analytic expression of Eq.~\ref{eq:formula} (red line), which neglects the $m_{e}$ terms, and the FANG calculation (black points), and the (blue histogram) of events generated by FANG.}

    \label{fig:dSdO-3MeV}
\end{figure}

Fig.~\ref{fig:dSdO-3GeV} and Fig. \ref{fig:dSdO-3MeV} compare the analytic expression of Eq.~\ref{eq:formula} (red line) with the FANG differential cross section calculation using point generation routine at specific points (black points), and the blue histogram of events generated by FANG using the user-defined routine. The user-defined routine allows the constraining vectors to be chosen anisotropically in the laboratory frame, provided the events are subsequently reweighted to ensure isotropy.
In this example, the vectors were generated with an angular distribution proportional to $1/\sin^{4}(\tfrac{\theta}{2})$. The events were then reweighted by a factor of $\sin^{4}(\tfrac{\theta}{2})$, which cancels the corresponding factor in $|\mathcal{M}|^{2}$. In this way, the variation of the event weights due to the strong angular dependence of the squared amplitude is removed, significantly reducing the statistical uncertainty in the weighted sum. In all cases the calculation was done by inserting $|\mathcal{M}|^{2}$ from Eq.~\ref{eq:emuAmp} into Eq.~\ref{eq:dsigmadO}.
 For high energies, such as a $3~\text{GeV}$ electron (Fig.~\ref{fig:dSdO-3GeV}), the analytical and FANG calculations coincide. At lower energies, such as a $3~\text{MeV}$ electron (Fig.~\ref{fig:dSdO-3MeV}), a divergence appears, as expected, since the analytical calculation neglects all terms proportional to the electron mass, while the Monte Carlo includes the full expression. \\
Another closely related example is elastic $ep \rightarrow ep$ scattering taken from \cite{eichmann_qcd_2020}, 
\begin{equation}
\label{eq:epAmp}
\overline{|\mathcal{M}|^2}_{ep\to ep}= \frac{e^4}{q^4} L^{\mu\nu} W_{\mu\nu} 
= \frac{16 \pi^2 \alpha^2}{\tau^2} 
\left(\frac{G_E^2 + \tau G_M^2}{1+\tau} (\lambda^2 + \tau^2 - \tau) + 2\tau^2 G_M^2\right)
\end{equation}
Here, the dimensionless variables $\tau$ and $\lambda$ are defined as  
\begin{align}
\tau &= -\frac{q^2}{4M^2}
&\qquad 
\lambda &= \frac{(p+p') \cdot (k+k')}{4M^2}
\end{align}
where $M$ is the nucleon mass and $q^2 \equiv -Q^2$ denotes the invariant four--momentum transfer squared. 
The proton four-momenta before and after scattering are denoted by $p$ and $p'$, respectively, while $k$ and $k'$ are the corresponding electron four-momenta.  

\begin{figure}[H]
    \centering
    \includegraphics[width=0.8\linewidth]{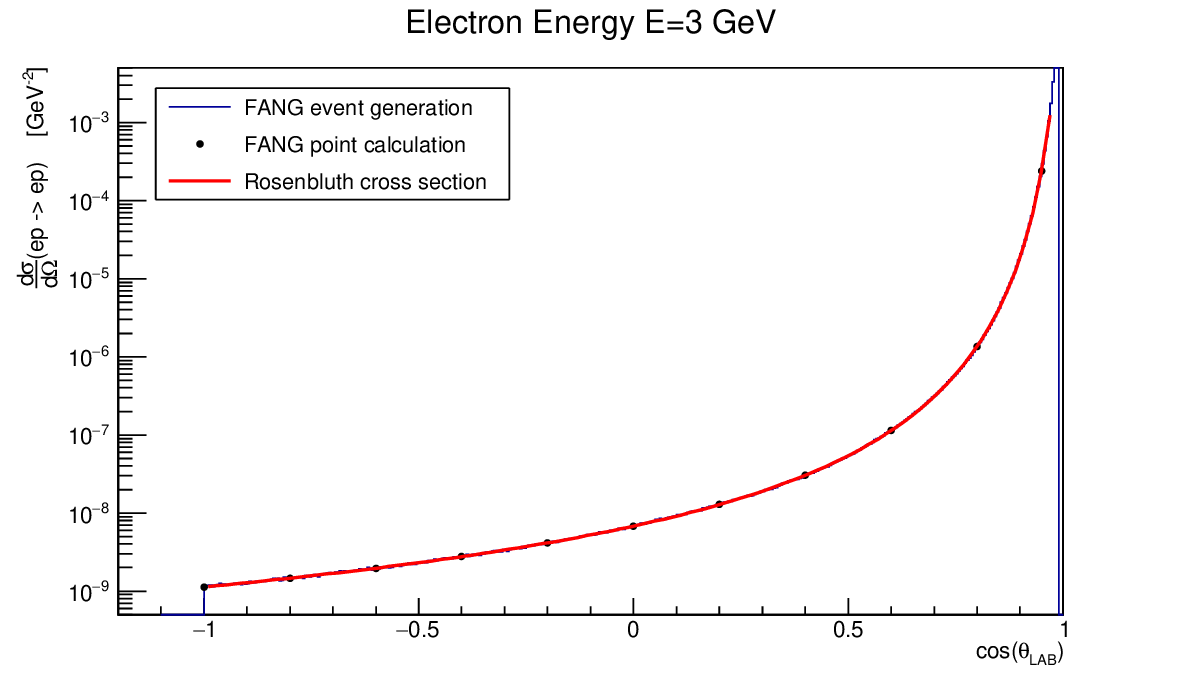}
\caption{Differential $e^{-}p$ elastic cross section, showing agreement between the Rosenbluth formula (Eq.~8.17 in~\cite{halzen_quark_2008}, red line) and the FANG calculation (black points), and the (blue histogram) of events generated by FANG.}

    \label{fig:Elastic-ep}
\end{figure}

The electric and magnetic form factors are taken in the standard dipole form,
\begin{equation}
G_E(Q^2) = \frac{G_M(Q^2)}{\mu} 
= \left(1+\frac{Q^2}{0.71 \,\text{GeV}^2}\right)^{-2},
\end{equation}
with $\mu = 2.793$ the proton magnetic moment.  
Based on the above definitions, the FANG calculation of the differential cross section at selected kinematic points can be benchmarked against the Rosenbluth formula \cite{rosenbluth_high_1950}, providing a direct test of the method.
The results of this comparison are shown in Fig.~\ref{fig:Elastic-ep}.
All three figures also show excellent agreement with the angular distributions of the generated events, displayed as histograms in Figs.~\ref{fig:dSdO-3GeV}--\ref{fig:Elastic-ep}.

\subsection{Testing FANG with Lepton–Nucleus Scattering Models}

GENIE \cite{andreopoulos_genie_2010} is a widely used event generator for interactions of neutrinos (or electrons) with all nuclear targets and flavors across a broad energy range, from approximately $1$ MeV up to $1$ PeV. In GENIE, the relativistic Fermi gas (RFG) nuclear model is applied to all processes. The most stringent tests of the RFG model come from electron scattering experiments \cite{papadopoulou_inclusive_2021}.

Within the GENIE framework, the electron energy transfer distribution in the laboratory frame, $\nu = E_e - E'_e$, in scattering at a fixed (large) angle is strongly correlated with the nucleon momentum distribution described by the RFG model. To obtain sufficient statistics from GENIE simulations restricted to, for example, electron scattering at $60 \pm 0.5^{\circ}$, it is necessary to impose a $Q^2$ cut. However, this cut in turn limits the maximum accessible energy transfer, preventing simulation of the full measured distribution. Moreover, after applying a $Q^2$ cut, the majority of simulated events correspond to forward-going electrons that do not contribute to the desired kinematics. As a result, the process becomes both inefficient and computationally expensive.  

In contrast, FANG can directly generate only those events where the electron scatters into the acceptance region (e.g., $60 \pm 0.5^{\circ}$), without requiring a $Q^2$ cut. This allows simulation of the full energy transfer distribution with much higher efficiency.

\subsubsection{Implementing the GENIE Physics Model in FANG}

The potential usefulness of FANG for this type of analysis was demonstrated by simulating electron scattering at $E_e = 0.961 \, \mathrm{GeV}$ on $^{12}\mathrm{C}$ at several fixed angles, and comparing the resulting energy transfer distributions from FANG with those from GENIE.  

In GENIE, electron–nucleus scattering is modeled as a two-step process. First, the electron scatters from a single nucleon or nucleon pair, as described by the RFG model, producing the outgoing electron and a primary hadronic system. In the second step, the primary hadronic system propagates through and interacts with the residual nucleus, producing the final hadronic state. Since the energy transfer is determined entirely in the first step, FANG was used here only to simulate this stage of the reaction.

For comparison, a GENIE sample of $2 \times 10^7$ events was generated for $0.961 \, \mathrm{GeV}$ electrons scattering on $^{12}\mathrm{C}$, with a $Q^2 > 0.1 \, \mathrm{GeV}^2$ cut. The nucleon momentum distribution and the primary hadronic mass distribution extracted from this GENIE output were then used as input to FANG.

As in GENIE, FANG events were generated in four separate channels:
\begin{itemize}
    \item Quasi-elastic (QE): the electron scatters from an initial nucleon, and the outgoing primary hadronic system is a proton or neutron.
    \item Resonance (RES): similar to QE, but the outgoing primary hadronic state is a nucleon resonance such as $\Delta(1232)$ or $N(1440)$.
    \item Meson-exchange currents (MEC): the electron scatters from a correlated nucleon pair, producing two outgoing nucleons, usually a proton and a neutron. This channel is subdominant and becomes negligible at large angles.
    \item Deep inelastic scattering (DIS): the electron scatters from a quark in the initial nucleon, and the outgoing hadronic system corresponds to the breakup of the nucleon.
\end{itemize}

Each channel has its own cross section. Extracting the exact squared matrix element $|\mathcal{M}|^2$ that reproduces the GENIE cross section for each channel is beyond the scope of this simple demonstration. Instead, the total number of events in FANG was normalized to the GENIE sample for each channel separately. The following simplified amplitudes were used: for the QE and RES channels, the same elastic scattering amplitude as Eq.~\ref{eq:epAmp} was employed with the appropriate four-momenta; the MEC channel was generated using isotropic phase space ($|\mathcal{M}|^2=1$); and for the DIS channel, the amplitude for $ep \rightarrow eX$ scattering was taken from Ref.~\cite{halzen_quark_2008}:
\begin{equation}
\label{eq:DISAmp}
\overline{|\mathcal{M}|^2}_{ep\rightarrow eX}=\frac{128\pi^2\alpha^2}{q^4} \left[4W_1 (k'\cdot k') +\frac{2W_2}{M^2} \left(2(k'\cdot p)(k\cdot p') -M^2 k' \cdot k  \right)\right]4\pi M,
\end{equation}
where $p$ is the four-momentum of the initial nucleon, $p'$ the four-momentum of the outgoing hadronic system, and $M$ its mass.  
For simplicity, the following relations are used: the structure functions are written as 
$\nu W_2 = F_2$ and $MW_1 = \tfrac{F_2}{2x}$, 
the Bjorken scaling variable is defined as $x = \tfrac{Q^2}{2M\nu}$, 
and the neutron structure function is approximated by 
$F^{en}_2 = F^{ep}_2 (1 - 0.75x)$.
where the proton structure function $F^{ep}_2(x,Q^2)$ is taken from the ALLM97 parametrization \cite{abramowicz_allm_2004,abramowicz_parametrization_1991}, a 23-parameter fit to world data.

\subsubsection{Comparison of FANG and GENIE Energy Transfer Results}

FANG generated $10^7$ events at each scattering angle, all constrained to lie within the $\pm 0.5^{\circ}$ detector acceptance, without applying any $Q^2$ cut. These were compared with the GENIE sample of $2 \times 10^7$ inclusive ($4\pi$) events generated with $Q^2 > 0.1 \, \mathrm{GeV}^2$.  

\begin{figure}[H]
    \centering
    \includegraphics[width=1\linewidth]{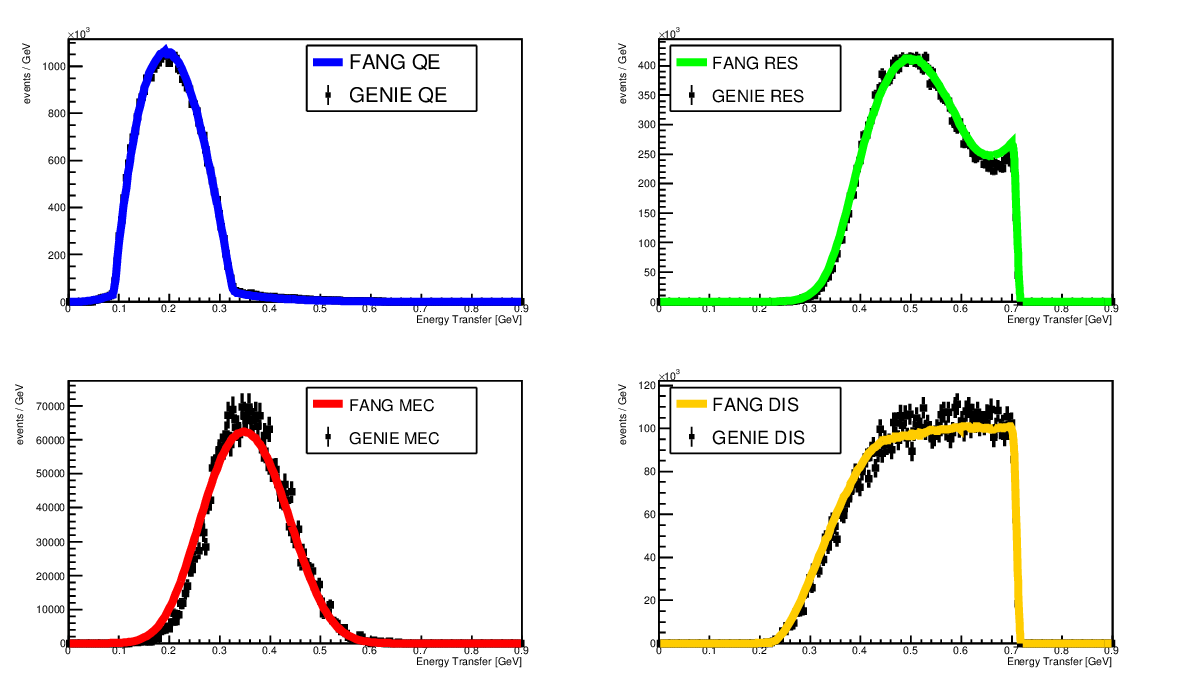}
    \caption{Comparison of FANG and GENIE energy transfer distributions for each channel separately, for electron scattering in the angular range $\theta=37.5\pm 0.5^{\circ}$.}
    \label{fig:FANG-GENIE37}
\end{figure}

For scattering in the angular range $\theta = 37.5 \pm 0.5^{\circ}$, Fig.~\ref{fig:FANG-GENIE37} shows excellent agreement between FANG and GENIE for the energy transfer distributions in each channel, with FANG providing much higher effective statistics. For direct comparison, the same $Q^2 > 0.1 \, \mathrm{GeV}^2$ cut was also applied to the FANG sample, reproducing the GENIE cutoff at large $\nu$.  

\begin{figure}[H]
    \centering
    \includegraphics[width=1\linewidth]{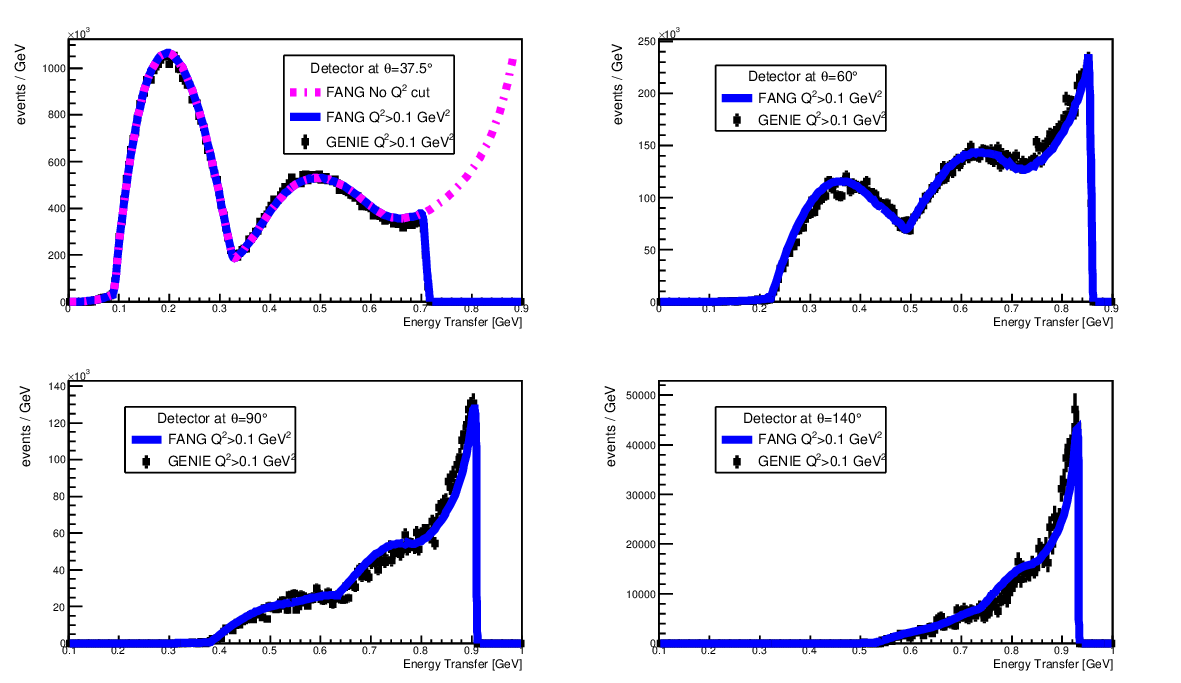}
    \caption{Comparison of FANG and GENIE energy transfer distributions at different angles. For $\theta=37.5\pm 0.5^{\circ}$, the FANG result without a $Q^2$ cut is also shown.}
    \label{fig:FANG-GENIE}
\end{figure}

Fig. \ref{fig:FANG-GENIE} shows further comparisons at different scattering angles. At $\theta=37.5\pm 0.5^{\circ}$ the advantage of FANG namely, the ability to generate the full energy transfer distribution without artificial cutoffs from a $Q^2$ requirement is clearly demonstrated.

For realistic physics analyses, however, the matrix element should be replaced with those corresponding to the exact cross section calculations implemented in GENIE, rather than the simplified approximations used here.

\section{Summary}
\label{sec:summary}
We have developed FANG (Focused Angular $n$-body event Generator), a Monte Carlo algorithm designed for efficient event generation in restricted Lorentz-invariant phase space. In contrast to conventional generators that sample the full solid angle, FANG directly imposes laboratory-frame angular constraints while preserving exact kinematics and correct LIPS weighting. This targeted approach eliminates the need for post-selection cuts, achieving orders-of-magnitude gains in computational efficiency for angular observables.

The algorithm has been validated against analytic results, standard generators (RAMBO, GENBOD), and established simulation frameworks (GENIE), showing excellent agreement across phase-space volumes, angular and energy distributions, and differential cross sections. In benchmark scenarios, FANG reproduces known analytic cross sections and provides significant efficiency improvements when simulating detector-relevant geometries where only small fractions of phase space contribute.

By combining exact kinematics with computational efficiency, FANG offers a versatile tool for differential cross section and decay-rate calculations in high-energy and nuclear physics. Its robustness makes it well suited for applications ranging from precision tests of the Standard Model to detector design and analysis in particle and nuclear experiments.

\acknowledgments

We thank Adi Ashkenazi and her research group at Tel Aviv University for their generous help and insightful discussions on the GENIE-related aspects of this work. We also thank Israel Mardor for helpful comments on an earlier version of this manuscript


\bibliographystyle{JHEP}
\bibliography{references}

\end{document}